# A NOTE ON THE DQ ANALYSIS OF ANISOTROPIC PLATES

WEN CHEN[*], WEIXING HE[**], AND TINGXIU ZHONG[*]

[*]Affiliation: Department of Mechanical Engineering, Shanghai Jiao Tong University, Shanghai 200030, P. R. China

[**]Affiliation: Department of Electrical Engineering, Jiangsu University of Science & Technology, Zhenjiang, Jiangsu 212013, P. R. China.

**Corresponding author**: Wen Chen (PhD)**, Mail address:** P. O. Box 9601BB, Shanghai Jiao Tong University, Shanghai 200030, P. R. China. **Fax**: 0086-021-62829425, **Tel**: 0086-021-62812679, **E-mail**: ctwang@sjtu.edu.cn

The total number of pages is 6.

# 1. INTRODUCTION

Recently, Bert, Wang and Striz [1, 2] applied the differential quadrature (DQ) and harmonic differential quadrature (HDQ) methods to analyze static and dynamic behaviors of anisotropic plates. Their studies showed that the methods were conceptually simple and computationally efficient in comparison to other numerical techniques. Based on some recent work by the present author [3, 4], the purpose of this note is to further simplify the formulation effort and improve computing efficiency in applying the DQ and HDQ methods for these cases.

# 2. APPROXIMATE FORMULAS IN MATRIX FORM AND REDUCTION COMPUTATIONS

The details about the DQ and HDQ methods see reference [1]. The only difference between DQ and HDQ methods is to choose different basis functions, namely, the former is based on the polynomials and the latter based on harmonic functions.

## 2.1 Approximate formulas in matrix form

Approximate formulas for partial derivatives of function w(x,y) in two-dimensional domain are given in matrix form by [3]

$$\frac{\partial^4 \hat{w}}{\partial x^4} = \overline{D}_x \hat{w}, \quad \frac{\partial^4 \hat{w}}{\partial x^3 \partial y} = \overline{C}_x \hat{w} \overline{A}_y^T, \quad \frac{\partial^4 \hat{w}}{\partial x^2 \partial y^2} = \overline{B}_x \hat{w} \overline{B}_y^T, \quad \frac{\partial^4 \hat{w}}{\partial x \partial y^3} = \overline{A}_x \hat{w} \overline{C}_y^T,$$
$$\frac{\partial^4 \hat{w}}{\partial y^4} = \hat{w} \overline{D}_y^T, \quad \frac{\partial^2 \hat{w}}{\partial x^2} = \overline{B}_x \hat{w}, \quad \frac{\partial^2 \hat{w}}{\partial y^2} = \hat{w} \overline{B}_y^T$$
(1)

where the unknown $\hat{w}$ is a n×m rectangular matrix rather than a vector as in references [1, 2], n and m is the number of inner grid points along x- and y- directions, respectively. $\overline{A}$, $\overline{B}$, $\overline{C}$ and $\overline{D}$ with subscripts x and y here stand for the DQ weighting coefficient matrices, modified by the respective boundary conditions using Wang and Bert's new approach [5], for the 1st, 2nd, 3rd and 4th order partial derivatives, respectively. The superscript T means the transpose of the matrices. It is noted that present DQ approximate formulas in matrix form can be easily extended to three-dimensional problems. The desired $\hat{w}$ in rectangular matrix form can be converted into the conventional vector form by the following Lemma 1.

**Lemma 1.** If $A \in C^{p \times m}$, $B \in C^{n \times q}$ and the unknown $X \in C^{m \times n}$, then

$$vec(AXB) = (A \otimes B^T)vec(X) \qquad (2)$$

where vec( ) is the vector-function of a rectangular matrix formed by stacking the rows of matrix into one long vector, $\otimes$ denotes the Kronecker product of matrices. To simplify the presentation, we define $vec(AXB) = AX\vec{B}$ and $vec(X) = \vec{X}$.

Corollary:

$$
\begin{aligned}
&1\ \ A\vec{X} = (A \otimes I_n)\vec{X} \\
&2\ \ X\vec{B} = (I_m \otimes B^T)\vec{X} \\
&3\ \ A\vec{X} + X\vec{B} = (A \otimes I_n + I_m \otimes B^T)\vec{X}
\end{aligned}
\qquad (3)
$$

where $I_n$ and $I_m$ are the unit matrix.

### 2.2 Centrosymmetric matrices and computing reduction

The weighting coefficient matrices in the DQ and HDQ methods were proven to be either centrosymmetric matrix for the derivatives of even order or skew centrosymmetric matrix for the derivatives of odd order if a grid spacing is symmetric [3, 4]. This is often seen in many situations such as equally spaced grid points or zeros of the Chebyshev and the Legendre polynomials. In the following we establish the notations on the centrosymmetric and skew centrosymmetric matrices first [3, 4, 6].

**Definition 1**. Let $V_{N \times N}$ and $NV_{N \times N}$ denote the set of N×N real centrosymmetric and skew centrosymmetric matrices, respectively, then $X=[x_{ij}] \in V_{N \times N}$ if and only if $x_{N+1-i,N+1-j} = x_{ij}$; and $X=[x_{ij}] \in NV_{N \times N}$ if and only if $x_{N+1-i,N+1-j} = -x_{ij}$.

**Lemma 2**. If $Q_1, Q_2 \in V_{N \times N}$; $Q_3, Q_4 \in NV_{N \times N}$, then $Q_1 Q_2 \in V_{N \times N}$, $Q_1+Q_2 \in V_{N \times N}$, $Q_3 Q_4 \in V_{N \times N}$

The following lemma 3 is presented on the Kronecker product of the centrosymmetric and skew centrosymmetric matrices. The proofs are straightforward and thus omitted for the sake of brevity.

**Lemma 3**. If $A_1$, $A_2 \in V_{N \times N}$ and $B_1$, $B_2 \in NV_{N \times N}$, then $A_1 \otimes A_2 \in V_{N \times N}$, $B_1 \otimes B_2 \in V_{N \times N}$, $A_1 \otimes B_1 \in NV_{N \times N}$.

The centrosymmetric and skew centrosymmetric matrices can be factorized into two smaller sub matrices in the calculation of their determinant, inverse and eigenmodes. Therefore, the respective computing effort and storage requirements can be reduced by 75 percent and 50 percent, respectively. The details on centrosymmetric and skew centrosymmetric matrices can be found in references [3, 4, 6].

### 3. ON ANISOTROPIC PLATES

The equation governing the behaviors of mid-plane symmetric laminated plates is given by [1]

$$\overline{D}_{11} w_{,xxxx} + 4\overline{D}_{16} w_{,xxxy} + 2(\overline{D}_{12} + \overline{D}_{66}) w_{,xxyy} + 4\overline{D}_{26} w_{,xyyy} + \overline{D}_{22} w_{,yyyy} \\ = q + \rho h \omega^2 w - N_x w_{,xx} - N_y w_{,yy} - 2N_{xy} W_{,xy}, \quad (4)$$

where $\overline{D}_{ij}$ are the plate stiffness, h is the total plate thickness, ρ is the density, w is the model deflections, q is the pressure only for deflection analysis, ω is the natural frequency only for free vibration analysis, $N_x$ and $N_y$ are uniform compression in-plane loads in the x- and y- directions for buckling analysis.

In terms of the present DQ and HDQ approximate formulas (1) with $\overline{N} = N_x = N_y$ and $N_{xy} = 0$, we have

$$\overline{D}_{11} \overline{D}_x \hat{w} + 4\overline{D}_{16} \beta \overline{C}_x \hat{w} \overline{A}_y^T + 2\beta^2 (\overline{D}_{12} + \overline{D}_{66}) \overline{B}_x \hat{w} \overline{B}_y^T + 4\overline{D}_{26} \beta^3 \overline{A}_x \hat{w} \overline{C}_y^T \\ + \overline{D}_{22} \beta^4 \hat{w} \overline{D}_y^T = qa^4 + \varpi^2 \hat{w} - \overline{N} a^2 (\overline{B}_x \hat{w} + \hat{w} \overline{B}_y^T), \quad (5)$$

where β=a/b denotes the aspect ratio, $\varpi^2 = \rho h a^4 \omega^2$. The relative boundary conditions have been taken into account in the formulation of weighting coefficient matrices, no additional equations are required.

Applying Lemma 1 and relative corollaries yields

$$\left[\overline{D}_{11}\left(\overline{D}_x \otimes I_y\right) + 4\overline{D}_{16}\beta\left(\overline{C}_x \otimes \overline{A}_y\right) + 2\beta^2\left(\overline{D}_{12} + \overline{D}_{66}\right)\left(\overline{B}_x \otimes \overline{B}_y\right) + 4\overline{D}_{26}\beta^3\left(\overline{A}_x \otimes \overline{C}_y\right) \right. \\ \left. + \overline{D}_{22}\beta^4\left(I_x \otimes \overline{D}_y\right)\right]\bar{w} = qa^4 + \varpi^2 \bar{w} - \overline{N}a^2\left(\overline{B}_x \otimes I_y + I_x \otimes \overline{B}_y\right)\bar{w} \quad (6)$$

The above formulation equation is equivalent to equation (13) in references [1, 2]. The present procedures obviously simplify formulation effort and are much easier for programming, and the resulting formulation has an explicit matrix form. It can be concluded that the approximate formulas in matrix form in the DQ and HDQ methods are much simpler, more compact and convenient to formulate partial differential operators in two-dimensional domain than the conventional ones in polynomial form presented by Civan and Sliepcevich [7].

For problems with symmetric boundary conditions such as the simply supported or clamped anisotropic plates as discussed in references [1, 2], it is straightforward that $\overline{A}_x$, $\overline{A}_y$, $\overline{C}_x$ and $\overline{C}_y$ are skew centrosymmetric matrix and $\overline{B}_x$, $\overline{B}_y$, $\overline{D}_x$ and $\overline{D}_y$ are centrosymmetric matrix when the uniform grid points or the zeros of the Chebyshev polynomials are used. According to Lemmas 2 and 3, the resulting coefficient matrix in the formulation equation (6) is a centrosymmetric matrix. Therefore, the reduction algorithm based on the factorization properties of centrosymmetric matrix [3, 4, 6] is applicable for the cases, namely, the computing effort and storage requirements are reduced to 75 percent and 50 percent as much as that in references [1, 2].

Bert et al. [1, 2] pointed out that the DQ and HDQ methods were very competitive technique to analyze static and dynamic behaviors of the anisotropic plates. The present work makes the methods more computationally efficient and easier to be used for these problems.